# Bridging cone-beam CT and MRI to stopping power ratio maps: a modality-agnostic Brownian bridge approach to support robust adaptive proton therapy

Yuhao Yan, Ph.D.[1,2], Qisheng He, Ph.D.[3*], Minglei Kang, Ph.D.[1], Behzad Hejrati, M.S.[4], Christian Hyde, M.D.[1], Ming Dong, Ph.D.[4], and Carri K. Glide-Hurst, Ph.D.[1,2]

1 Department of Radiation Medicine, University of Wisconsin-Madison, Madison, WI, USA
2 Department of Medical Physics, University of Wisconsin-Madison, Madison, WI, USA
3 Henry Ford Health, Detroit, MI, USA
4 Department of Computer Science, Wayne State University, Detroit, MI, USA
* Qisheng He, Ph.D. was affiliated with Department of Computer Science, Wayne State University, Detroit, MI, USA at the time this work was conducted.

**CORRESPONDENCE**
Carri Glide-Hurst, Ph.D., Department of Radiation Medicine, University of Wisconsin-Madison, 600 Highland Avenue, Madison, WI 53792.
Email: glidehurst@wisc.edu

Yuhao Yan, Ph.D., Department of Radiation Medicine, University of Wisconsin-Madison, 600 Highland Avenue, Madison, WI 53792.
Email: yyan228@wisc.edu

**ACKNOWLEDGEMENT**
The authors thank Ying Feng for developing the proton therapy plans. They thank Alex Kuo, M.S. for data screening and initial preprocessing. They also thank Dr. Jessica Miller, Ph.D. for supplying the Advanced Electron Density Phantom.

**FUNDING INFORMATION**
Work reported in this publication was supported in part by the National Cancer Institute of the National Institutes of Health under award number R01HL153720 (PI: Carri Glide-Hurst). The content is solely the responsibility of the authors and does not necessarily represent the official views of the National Institutes of Health. Research reported was supported in part by Nvidia Academic Grant and Lambda Inc.

**CONFLICT OF INTEREST STATEMENT**

Carri Glide-Hurst reports research collaborations with RaySearch Laboratories, Leo Cancer Care, Inc., and GE Healthcare (PI: Carri Glide-Hurst). Yuhao Yan, Qisheng He, Minglei Kang, Behzad Hejrati, Christian Hyde and Ming Dong report no conflict of interest.

**DATA AVAILABILITY STATEMENT**

The data cannot be made publicly available upon publication because they contain sensitive personal information. The data that support the findings of this study are available upon reasonable request from the authors.

## Abstract

**Objective:** Adaptive proton therapy (APT) relies on routine offline CT simulations for updated anatomy. To facilitate APT, we present a high-fidelity, modality-agnostic regularized Brownian bridge (rBBrg) framework to deterministically predict stopping power ratio (SPR) maps from cone-beam CTs (CBCTs) or MRIs.

**Approach:** rBBrg consisted of (1) generation mapping where a Brownian bridge diffusion model predicted SPR from input and (2) reconstruction mapping where a conditional generative adversarial network reconstructed input from predicted SPR. Class embedding was adopted for modality conditioning. Efficient, deterministic prediction was achieved via one-step sampling. Unimodality rBBrg and residual network (ResNet) were implemented for comparison. Matched-pair planning CT-CBCT (n=37) and planning CT-MRI (n=21) from head-and-neck cancer patients were evaluated. CBCTs and MRIs were deformably registered to CTs. A calibration phantom was scanned to establish Hounsfield look-up table for ground truth SPR. Predicted SPR maps were evaluated via mean absolute error (MAE), peak-signal-to-noise ratio and structural similarity between ground truth and predictions along with demonstration of dosimetric performance.

**Main results:** For CBCT-to-SPR synthesis, most quantitative metrics were comparable across models ($p \geq 0.05$). Modality-agnostic rBBrg outperformed ResNet in $MAE_{bone}$ (0.059±0.008 vs 0.067±0.008, Δ=11%, $p<0.05$) and showed slightly lower $MAE_{external}$ than ResNet (0.038±0.004 vs 0.040±0.004, $p=0.05$). Modality-agnostic and unimodality rBBrg were comparable ($p>0.05$). Qualitatively, both rBBrg models better preserved anatomical fidelity than ResNet. For MRI-to-SPR synthesis, quantitative and qualitative performance was overall comparable across models ($p>0.05$). Modality-agnostic rBBrg obtained $MAE_{external}$=0.057±0.005. Dose calculation comparisons between real and CBCT-predicted SPR were higher in modality-agnostic rBBrg predictions compared to ResNet with gamma pass rate=98.8%/99.3% vs 94.5%/96.9% at 2 mm/2%, respectively.

**Significance:** A novel modality-agnostic rBBrg model was developed to generate high-fidelity, deterministic SPR from CBCT or MRI to support efficient and flexible APT workflow. Initial feasibility has been established.

## Introduction

Proton therapy (PT) has demonstrated well-established dosimetric benefits for head and neck cancer (HNC) patients where it better spares multiple radiosensitive organs at risk (OARs) in complex anatomy over photon-based radiation therapy (RT) (Lillo *et al* 2024), thus leading to lower toxicities and improved quality of life (Frank *et al* 2026). A recent prospective, randomized phase III trial comparing Stage III-IV oropharyngeal cancer patients treated with intensity modulated proton therapy (IMPT) versus intensity modulated radiation therapy (IMRT) found improved 5-year overall survival rate, reduced grade≥3 toxicity including dysphagia and xerostomia, and less gastrostomy tube dependence for those treated with IMPT (Frank *et al* 2026).

Yet, the physical advantage of the finite range in PT makes it highly sensitive to daily anatomical variations, necessitating treatment adaptation especially for HNC patents (Evans *et al* 2020). Literature report that ~30-60% HNC patients treated by PT required adaptation at least once during the treatment course (Evans *et al* 2020, Stanforth *et al* 2022). In practice, quality assurance CTs (QACTs) are routinely acquired to verify treatment plans on the updated anatomy and assess the need of adaptation (Evans *et al* 2020), often weekly, however this is resource intensive and introduces additional imaging dose to patients, and the triggered offline adaptation could be delayed and not necessarily accommodate patients' anatomy of the day when delivered. Thus, the QACT pipeline remains a significant bottleneck in adaptive proton therapy (APT) workflow (Reiners *et al* 2023).

On-board cone beam CT (CBCT) is becoming increasingly available that reflects 3D volumes of patients' daily anatomy in the treatment position, which is ideal for online APT (Reiners *et al* 2023).

However, CBCT image quality is often compromised by scattering and beam hardening artifacts, thereby hindering direct, accurate proton dose calculations on CBCT (Park *et al* 2015). Enabling CBCT-based dose monitoring facilitates more efficient determination on treatment adaptation based on up-to-date patient anatomy, eliminates the need of QACT and associated uncertainties of patient setup and image registration between CT and CBCT, and streamlines the clinical workflow (Reiners *et al* 2023). To address this need, deep learning (DL)-based approaches have emerged to generate synthetic CT from CBCT, then the synthetic CT is used for proton dose calculation with intermediate steps of converting CT number to SPR (Pang *et al* 2023). Early solutions adopted convolutional neural network (CNN)-based or generative adversarial network (GAN)-based methods (Pang *et al* 2023), however these methods may yield blurred results (Pang *et al* 2023), fail to preserve image contrast (Zhang *et al* 2023), and not address changing organ shapes (de Hond *et al* 2023), which may undermine their ability for precise proton dose calculations and adaptive therapy.

MRI is playing an increasingly important role in radiation oncology departments in the form of MR-simulators (Glide-Hurst *et al* 2021) and MRI-guided radiation therapy in photon (Mulder *et al* 2022) and proton settings (Horst *et al* 2025). However, at present, MRI cannot directly provide the physical quantity, stopping power ratio (SPR), necessary for accurate proton dose calculations. Integrating high quality MRI-derived SPR maps into PT would unify high-fidelity target definition with accurate range estimation, while enabling functional imaging–guided response assessment and sparing of OARs. Together, this lays the groundwork for biologically informed optimization, where dose, range, linear energy transfer (LET), and tissue function can be co-optimized to personalize treatment beyond purely geometric planning. Similarly, CNN and GAN-based solutions have been investigated to generate synthetic CT from MRI followed by mapping SPR from synthetic CT for proton dose calculation (Thummerer *et al* 2020). A distinct challenge in MRI-based CT synthesis is that MRI signal intensity does not quantitatively correlate to physical properties such as electron density and SPR. A common failure mode is that both air and bone yield little signal in MRI, thus resultant generated synthetic CTs often suffer from misclassification of bone or air especially at complex tissue interfaces (Singhrao *et al* 2023), which may propagate to proton range errors. Other MR-specific factors, such as intensity nonuniformity and variability across MR scanners or protocols, also challenge the accuracy and generalizability of the model (Villegas *et al* 2024).

Diffusion model-based solutions are emerging which have demonstrated superior performance in image synthesis tasks such as preserving anatomical fidelity better over conventional DL methods (Peng *et al* 2024, Pan *et al* 2024), yet the model prediction is time consuming, making them unfit for rapid APT workflows, and the inherent stochasticity of the predictions requires careful evaluation (Peng *et al* 2024, Pan *et al* 2024). A deterministic solution with clinically acceptable inference time would be favored, for example approaching the inference time of conventional DL methods (<10 seconds per volume (Emami *et al* 2018)) to better fit within rapid adaptation workflows, which can be as short as approximately 20 minutes if online (Blumenfeld *et al* 2024). In addition, most models are implemented for a single input modality, whereas a modality-agnostic strategy would be beneficial to enable more flexibility for clinical implementation (Wongtrakool *et al* 2025). Furthermore, most solutions address CT synthesis, while CT number calibration to SPR is specific to imaging protocols and scanners (Peters *et al* 2023), thus training synthetic CT models with CT datasets acquired with different protocols on different scanners would potentially introduce ambiguity to CT number-SPR calibration. To address the above limitations, we developed and validated a novel high-fidelity modality-agnostic regularized Brownian Bridge (rBBrg) diffusion model framework that directly predicts SPR maps from arbitrary inputs of CBCT or MRI datasets. Our overarching goal is to enable high-fidelity and deterministic predictions that are suitable for rapid clinical HNC APT workflows regardless of input image modality.

# Methods

## Data

Fifty-one HNC patients (39 M/12 F, age: 41-86 years) treated between June 2024 and April 2026 at three network locations of our institution (n=31/19/1) were retrospectively collected. Patients were treated using volumetric modulated arc therapy (VMAT) to 20-82.5 Gy in 5-54 fractions, including 4 patients treated with hypofractionation and 1 with hyperfractionation. CT simulation was performed on Siemens (Siemens Healthineers, Erlangen, Germany) SOMATOM X.ceed (n=29) or Definition Edge (n=22) across two network locations. Post-contrast single energy CTs (SECTs) were acquired with 120 kVp, reconstruction kernel of Qr40s (vendor-defined regular quantitative kernel with standard scan mode) on X.ceed and Br38f (vendor-defined regular body kernel with fast scan mode) on Definition Edge, 500-650 mm reconstruction diameter, 0.98-1.27 $mm^2$ in-plane resolution, and 1-2 mm slice thickness. Siemens iMAR (Iterative Metal Artifact Reduction) was applied during reconstruction. First fraction TrueBeam (Varian, Palo Alto, CA) CBCTs, occurring 16.3±3.4 days (range: [10, 24] days) after CT simulation, for 37 patients of the cohort were collected with 125 kVp, Ram-Lak reconstruction kernel, 465-485 mm reconstruction diameter, 0.91-0.95 $mm^2$ in-plane resolution, and 2 mm slice thickness. Twenty-one of the patients (14 M/7 F) also underwent MR simulation in the treatment position (immobilization and custom head rest) on 1.5T GE (GE Healthcare, Waukesha, WI) SIGNA Artist MRI scanners with an interval of 1.6±2.6 days (range: [0, 9] days) from CT simulation. Pre-contrast T1 CUBE was acquired and reconstructed in the sagittal plane with 90º flip angle, 500 ms repetition time, ~24 ms echo time, 24 echo train length, 0.47-0.61 $mm^2$ in-plane resolution, and 1.2 mm slice thickness. Vendor-provided deep CNN-based reconstruction enhancement (AIR™ Recon DL, GE Healthcare) was adopted with medium signal-to-noise ratio improvement (strength=0.50) to reduce noise, artifacts and improve edge sharpness.

**Hounsfield Look-up Table (HLUT) specifications**

To calibrate CT numbers to SPR, HLUTs were derived for each CT scanner separately. An Advanced Electron Density Phantom (Sun Nuclear, Melbourne, FL) including tissue-mimicking inserts (lung, adipose, soft tissue and bone) with known density and chemical compositions was scanned on both CT scanners using protocols approaching the corresponding clinical protocols (120 kVp, reconstruction kernel = Qr40/Br38s on X.ceed/ Definition Edge, respectively, 500 mm reconstruction diameter = 500 mm, slice thickness = 2/3 mm on X.ceed/ Definition Edge, respectively, no iMAR). CT number of each insert was extracted using a Python algorithm developed based on Pylinac. Using the freely available code (Peters *et al* 2023), the SPR of each insert was calculated following Bethe equation with mean excitation energies from ICRU-49 (Berger MJ *et al* 1993). Combining tabulated human tissue data (Woodard and White 1986), piecewise linear regression was performed between CT numbers and SPRs to derive the final HLUT, grouped by four tissue types (lung, adipose, soft tissue and bone).

**Image preprocessing**

CBCT and MRI were deformably registered to SECT in MIM Maestro v7.4.2 (MIM software, Cleveland, OH) using multimodality profile with smoothness factor=0.5, then resampled to the SECT resolution as the common frame of reference. For CBCT, training data included slices extending superiorly to the brainstem and inferiorly to the apex of lung or the inferior extent of PTV if lower. Slices where anatomy was truncated or erroneously deformed due to mismatched imaging field-of-views (FOVs) were excluded. For MRI, shoulder was always truncated due to acquisition in sagittal plane with limited FOV, thus only slices with complete anatomy were included extending superiorly to the brainstem and inferiorly to right above the shoulder. All images were subsequently resampled to 1.5 $mm^2$ in-plane resolution and 2 mm slice thickness, then cropped to 352-by-224 matrix for model training. External contours were generated using vendor-provided "whole body" contouring algorithm (MIM) for SECT and CBCT and using Otsu thresholding followed by morphological operation for MRI. To mask out non-anatomical devices such as the couch and external patient supports, union masks were generated for CBCT-SECT and MRI-SECT pairs. MRI was further standardized applying N4 bias field correction followed by Nyul histogram matching (Nyúl *et al* 2000). SECT was mapped to SPR maps using established scanner-specific HLUTs. Finally, all images (SPR maps, CBCT, and T1-w MRI) were normalized to [0, 1] for model training. Specifically, SPR maps for CT were normalized from [0.0011, 2.6617] corresponding to [-1000, 3095] HU, CBCT was normalized from [-1000, 3095] gray value, and MRI was normalized from [1, 99] percentile of pixel intensity in foreground.

**High fidelity regularized Brownian bridge diffusion model (rBBrg)**

***Prior work in diffusion models***

Briefly, diffusion models learn to convert a known distribution (e.g., Gaussian noise) to the target data distribution, originally inspired by nonequilibrium thermodynamics (Sohl-Dickstein *et al* 2015). Diffusion models generally consist of two processes: a forward process which gradually adds noise to the data until it approaches a known noise distribution, and a reverse process where the model learns to reverse the forward process to generate samples of data from the noise (Sohl-Dickstein *et al* 2015). One of the foundational works in diffusion-based approaches is the denoising diffusion probabilistic model (DDPM) initially proposed as a generative model (Ho *et al* 2020). In DDPM, the forward process is a predefined, discrete-time Markov chain that step-by-step adds Gaussian noise to the data until the corrupted data is approximately distributed according to the standard Gaussian distribution, and during the reverse process, the model learns a parameterized Markov chain in the reverse direction to generate data from Gaussian noise. For image-to-image translation tasks, it has been investigated to condition DDPM on the source images, for example by directly concatenating the source image as an additional input channel to the model, which have achieved promising performance (Peng *et al* 2024).

Song *et al* proposed a unified framework for score-based generative modeling based on stochastic differential equations (SDEs), generalizing previous diffusion-based solutions including DDPM and guiding development of SDE-based diffusion models (Song *et al* 2020). Under the framework, "score" refers to the score function, which is the time-dependent gradient of the log probability density with respect to the perturbed data. Such gradient points to the direction that the log probability density of the perturbed data distribution increases the most. The model learns to estimate the score function which is then used to generate data from noise through the reverse-time SDE, defined below.

For notation, the data (or an image) is represented as $\mathbf{X}_t \in \mathbb{R}^{\boldsymbol{n}}$, where the subscript $t \in [0, \mathrm{T}]$ is a continuous time variable representing the state of the data during the diffusion process. More specifically, $\mathbf{X}_0$ represents a sample from the target data distribution, $\mathbf{X}_t$ represents the sample $\mathbf{X}_0$ perturbed with noise at diffusion time $t$, and $\mathbf{X}_\mathrm{T}$ represents the noisy sample at the final diffusion time $\mathrm{T}$, whose distribution conventionally approaches a simple known distribution such as Gaussian noise in DDPM. Song *et al* constructed the forward process, i.e., gradually perturbing data with noise, following the SDE (Song *et al* 2020):

$$\mathrm{d}\mathbf{X}_t = \mathbf{F}(\mathbf{X}_t, t)\mathrm{d}t + \mathrm{G}(t)\mathrm{d}\mathbf{W}_t$$

$\mathbf{W}_t \in \mathbb{R}^{\boldsymbol{n}}$ is a standard Wiener process (also known as Brownian motion). $\mathbf{F}(\cdot, \mathrm{t}): \mathbb{R}^{\boldsymbol{n}} \rightarrow \mathbb{R}^{\boldsymbol{n}}$ is the drift coefficient of $\mathbf{X}_t$. It is the deterministic component of the equation describing the average change of the stochastic process. $\mathrm{G}(\cdot): \mathbb{R} \rightarrow \mathbb{R}$ is the diffusion coefficient describing the magnitude of the noise perturbation. The forward process is often predefined, i.e., $\mathbf{F}(\mathbf{X}_t, t)$ and $\mathrm{G}(t)$ are known. The reverse process that generates data from the noise follows the reverse-time SDE:

$$\mathrm{d}\mathbf{X}_t = [\mathbf{F}(\mathbf{X}_t, \mathrm{t}) - \mathrm{G}(t)^2 \nabla_{\mathbf{x}} \log \mathrm{p}(\mathbf{X}_t, t)]\mathrm{d}t + \mathrm{G}(t)\mathrm{d}\overline{\mathbf{W}}_t$$

where $\mathrm{p}(\mathbf{X}_t, t)$ is the probability density function of $\mathbf{X}_t$, and $\overline{\mathbf{W}}_t$ is a standard Wiener process in the reverse direction of time (i.e., T to 0). The model is thus trained to estimate the *score* function $\nabla_{\mathbf{x}} \log \mathrm{p}(\mathbf{X}_t, t)$ at each time $t$ to generate samples of data from noise via the reverse-time SDE.

***Brownian bridge diffusion model***

Distinct from the conventional diffusion-based method such as conditional DDPM which learns to sample target images from Gaussian noises conditioned on source images, Brownian bridge diffusion models learn the direct translation between two image domains $A$ and $B$. Wang *et al* constructed a generalized framework of score-based Brownian bridge directly from SDEs (Wang *et al* 2024). For continuous $t \in [0, 1]$, the forward process is defined to start from a data sample $\mathbf{X}(t = 0) = \mathbf{X}_0 \in \boldsymbol{B}$ corresponding to the target domain of the overall task (e.g., SPR) and end at the corresponding data sample $\mathbf{X}(t = 1) = \mathbf{X}_1 \in \boldsymbol{A}$ in the source domain (e.g., MRI), thus forming a bridge between two fixed ends, as demonstrated in Figure 1. Specifically, they constructed the forward SDE for Brownian bridge as:

$$\mathrm{d}\mathbf{X}_t = -\frac{\mathbf{F}(\mathbf{X}_t)}{1-t}\mathrm{d}t + \mathbf{G}(\mathbf{X}_t, t)\mathrm{d}\mathbf{W}_t = -\frac{\mathbf{X}_t - \mathbf{X}_1}{1-t}\mathrm{d}t + \lambda\sqrt{1-t}\mathrm{d}\mathbf{W}_t$$

$$\mathbf{X}(t=0) = \mathbf{X}_0$$

where $\mathbf{F}(\mathbf{X}_t) = \mathbf{X}_t - \mathbf{X}_1$ and $\mathbf{G}(\mathbf{X}_t, t) = \lambda\sqrt{1-t}$ are the drift and diffusion coefficients of $\mathbf{X}_t$, respectively. $\lambda > 0$ is a hyperparameter, set as 2 in our work. Following the theorem proposed by Wang *et al* (Wang *et al* 2024), the forward process can be explicitly expressed as:

$$\mathbf{X}_t = (1-t)\mathbf{X}_0 + t\mathbf{X}_1 + \lambda(1-t)\int_0^t \frac{\mathrm{d}\mathbf{W}_\tau}{\sqrt{1-\tau}}$$

which gives the marginal distribution of the forward process given $\mathbf{X}_0$ and $\mathbf{X}_1$:

$$\mathbf{X}_t \sim \mathcal{N}\left((1-t)\mathbf{X}_0 + t\mathbf{X}_{1,}\, \lambda^2(1-t)^2\int_0^t \frac{\mathrm{d}\tau}{1-\tau} I\right)$$

Using the predefined $\mathbf{F}(\mathbf{X}_t)$ and $\mathrm{G}(\mathbf{X}_t, t)$, the reverse-time SDE can be expressed as:

$$\mathrm{d}\mathbf{X}_t = -(\frac{\mathbf{X}_t - \mathbf{X}_1}{1-t} + \lambda^2(1-t)\nabla_{\mathbf{x}}\log \mathrm{p}(\mathbf{X}_t, t))\mathrm{d}t + \lambda\sqrt{1-t}\mathrm{d}\overline{\mathbf{W}}_t$$

$$\mathbf{X}(t=1) = \mathbf{X}_1$$

For practical model training, above continuous formula needs discretization. For discrete time step with a total number of time steps T, note $\mathbf{X}(t=0) = \mathbf{X}_0, \mathbf{X}(t=\mathrm{T}) = \mathbf{X}_\mathrm{T}, \boldsymbol{\epsilon} \sim \mathcal{N}(0, \mathbf{I})$ representing the Gaussian noise, and the forward process can be discretized as:

$$\mathbf{X}_t = \left(1 - \frac{t}{\mathrm{T}}\right)\mathbf{X}_0 + \frac{t}{\mathrm{T}}\mathbf{X}_\mathrm{T} + \mathrm{B}(t)\boldsymbol{\epsilon}$$

$$\text{where } \mathrm{B}(t) = \begin{cases} \lambda\left(1 - \frac{t}{\mathrm{T}}\right)\sqrt{\ln(\frac{1}{1-\frac{t}{\mathrm{T}}})} & if\ t = 1, .., \mathrm{T}-1 \\ 0 & if\ t = \mathrm{T} \end{cases}$$

During the reverse process, $\mathbf{X}_0$, which corresponds to the target image (e.g., SPR), is unknown. The Brownian bridge model $\boldsymbol{\epsilon_\theta}$ is trained to learn the difference between $\mathbf{X}_t$ and $\mathbf{X}_0$:

$$L(\theta) = \nabla_\theta \|\mathbf{X}_t - \mathbf{X}_0 - \boldsymbol{\epsilon_\theta}(\mathbf{X}_t, t)\| = \nabla_\theta \left\| \frac{t}{\mathrm{T}}(\mathbf{X}_\mathrm{T} - \mathbf{X}_0) + \mathrm{B}(t)\boldsymbol{\epsilon} - \boldsymbol{\epsilon_\theta}(\mathbf{X}_t, t) \right\|$$

Using $\widehat{\mathbf{X}}_0 = \mathbf{X}_t - \boldsymbol{\epsilon_\theta}(\mathbf{X}_t, t)$ in lieu of $\mathbf{X}_0$ together with the corresponding marginal distribution of $\mathbf{X}_t$, the intermediate steps of the reverse process can be discretized as

$$\mathbf{X}_{t-1} = C_{xt}\mathbf{X}_\mathrm{t} - C_{yt}\mathbf{X}_\mathrm{T} - C_{\epsilon t}\boldsymbol{\epsilon_\theta}(\mathbf{X}_t, t) - C_{zt}\boldsymbol{z}, \quad t = T, \ldots, 2$$

where $\boldsymbol{z} \sim \mathcal{N}(0, \mathbf{I})$, and expression of coefficients $C_{xt}$, $C_{yt}$, $C_{\epsilon t}$ and $C_{zt}$ are listed in the supplement materials. For t=1, $\mathbf{X}_0 = \mathbf{X}_1 - \boldsymbol{\epsilon_\theta}(\mathbf{X}_1, 1)$. Together, the reverse process progressively refines the intermediate sample toward the target image $\mathbf{X}_0$ to obtain a high-quality final prediction.

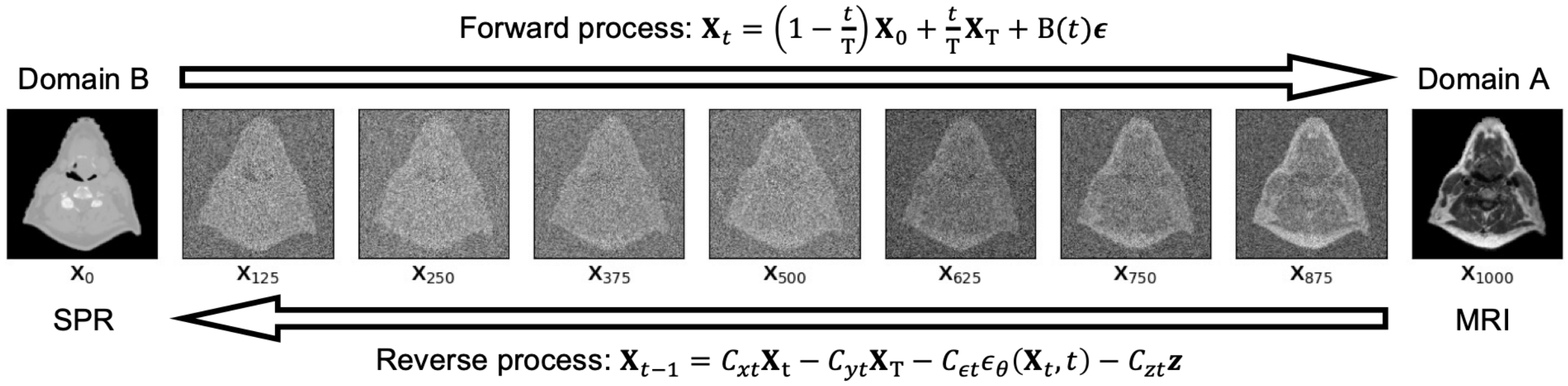

***Figure 1.*** *Schematic of Brownian bridge diffusion processes, taking translating from MRI (domain A) to stopping power ratio (SPR) map (domain B) for an example. The noising processes are tied to two fixed ends.*

***Regularized Brownian bridge diffusion model (rBBrg)***

The proposed regularized Brownian bridge diffusion model (rBBrg) consists of a generation mapping and a reconstruction mapping as demonstrated in Figure 2 (He *et al* 2025). The generation mapping is a conditional Brownian bridge which predicts target images (e.g., SPR) from the source images (e.g., CBCT/MRI), where the model adopts previously described Brownian bridge framework while conditioned on the source image via direct concatenation as an additional input channel, such that the training objective is

$$L_{diff} = \|(\mathbf{X}_t - \mathbf{X}_0) - \boldsymbol{\epsilon}_{\boldsymbol{\theta}}(\mathbf{X}_t, \mathbf{X}_{\mathrm{T}}, t)\|$$

Thus, the predicted image (SPR) is calculated as $\widehat{\mathbf{X}}_0^t = \mathbf{X}_t - \boldsymbol{\epsilon}_{\boldsymbol{\theta}}(\mathbf{X}_t, \mathbf{X}_{\mathrm{T}}, t)$. The reconstruction mapping is a conditional generative adversarial network (cGAN) consisting of a generator $G$ and a discriminator $D$. It reconstructs the source images (CBCT/MRI) from the predicted images (SPR): $\widehat{\mathbf{X}}_{\mathrm{T}}^t = G(\widehat{\mathbf{X}}_0^t)$. At each time step, we enforce the fidelity of reconstruction by

$$L_{rec} = \left\|\mathbf{X}_{\mathrm{T}} - \widehat{\mathbf{X}}_{\mathrm{T}}^t\right\|$$

as well as incorporating adversarial learning as follows

$$L_{adv} = \mathbb{E}_{\widehat{\mathbf{X}}_{\mathrm{T}}^t \sim \mathrm{p}(\widehat{\mathbf{X}}_{\mathrm{T}}^t)}\left[\log\left(1 - \sigma\left(D\left(\widehat{\mathbf{X}}_{\mathrm{T}}^t\right)\right)\right)\right] + \mathbb{E}_{\mathbf{X}_{\mathrm{T}} \sim \mathrm{p}(\mathbf{X}_{\mathrm{T}})}[\log \sigma(D(\mathbf{X}_{\mathrm{T}}))]$$

where $\sigma(\cdot)$ denotes the sigmoid function. The final training objective is

$$L = L_{diff} + \lambda_{rec} L_{rec} + L_{adv}$$

The reinforcement at each time step enforces the sampling trajectory of proposed rBBrg condensed to the central one with a much lower variance compared to other solutions, thus enabling high-quality one-step sampling which is efficient and deterministic:

$$\widehat{\mathbf{X}}_0^{\mathrm{T}} = \mathbf{X}_{\mathrm{T}} - \boldsymbol{\epsilon}_{\boldsymbol{\theta}}(\mathbf{X}_{\mathrm{T}}, \mathbf{X}_{\mathrm{T}}, \mathrm{T})$$

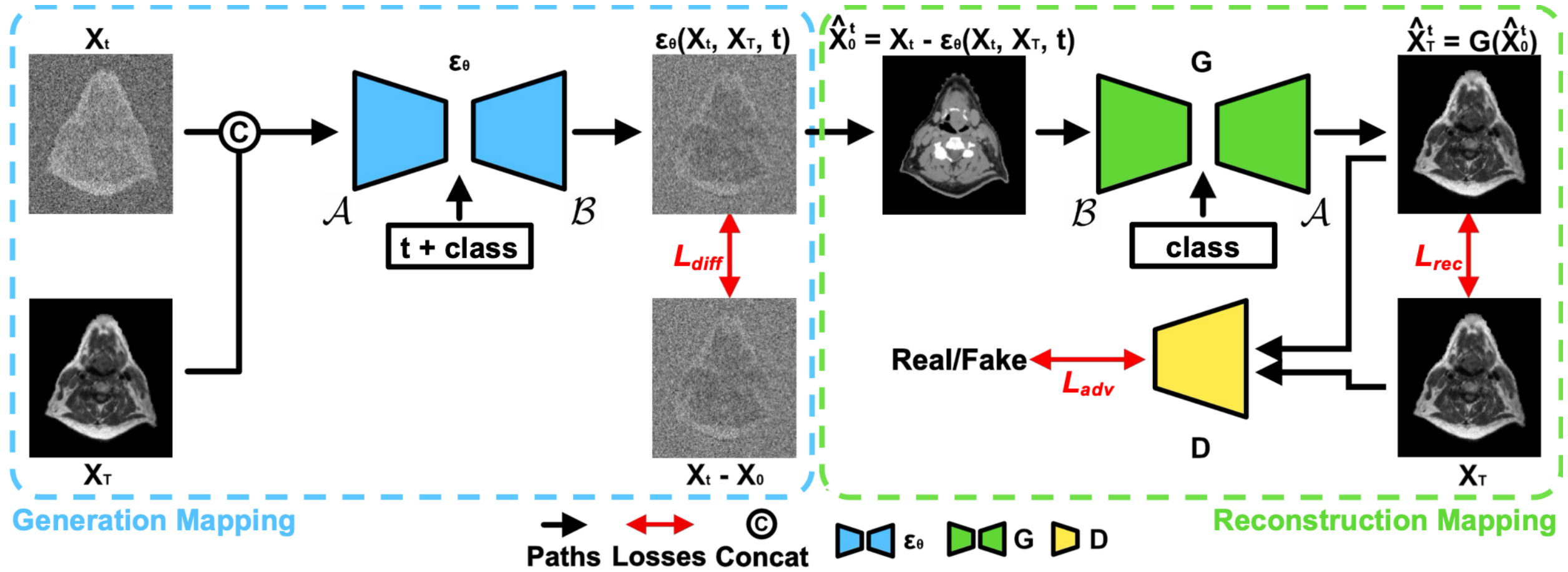

***Figure 2.*** *Architecture of the regularized Brownian bridge diffusion model, consisting of (left) a generation mapping where a conditional Brownian bridge predicts stopping power ratio (SPR) maps (domain B) from source images (CBCT/MRI, domain A), and (right) a reconstruction mapping where a conditional generative adversarial network reconstructs source images from the predicted SPR maps providing reinforcement on prediction fidelity and adversarial learning. Class embedding indicating the source image modality is incorporated to yield a modality-agnostic model. For the unimodality model, class embedding is omitted.*

***Implementation***

For the unimodality model, the Brownian bridge in generation mapping utilized a 2D U-net with residual blocks and attention blocks (Nichol and Dhariwal 2021). Time embedding was incorporated following FiLM-like (Perez *et al* 2018) approach. The cGAN in reconstruction mapping implemented the same U-net architecture without time embedding as the generator and PatchGAN (Isola *et al* 2017) as discriminator. For the modality-agnostic model, class embedding based on input modality was additionally incorporated for both U-net, and two modality-specific discriminators were used for CBCT/T1-w MRI respectively.

For all models, the following parameters were used: total number of time steps T = 1000, λ in forward process = 2, $\lambda_{rec}$ = 0.2 and 0.02 for the CBCT and T1-weighted MRI models, respectively. For the modality-agnostic model, $\lambda_{rec}$ was dynamic corresponding to the input modality. Brownian bridge and the generator in reconstruction mapping were optimized using Adam optimizer with learning rate = 2e-5, betas = 0.9 and 0.999, and eps = 1e-8. The discriminator was optimized using Adam optimizer with the same parameters except learning rate = 2e-4. Training epochs were set to 200. Batch size was set at 2. Models were implemented using PyTorch 2.5.1 on a 140 GB Nvidia H200 GPU. To comprehensively evaluate the models, five-fold cross-validation was implemented (Bradshaw *et al* 2023). Modality-agnostic model and unimodality models used the same dataset split. Data augmentation was implemented including translation (±20 pixels along each direction), rotation (±6°) and scaling (±15%).

**Quantitative evaluation**

Synthetic SPR maps were restored to the physical dynamic range ([0.0011, 2.6617]) for evaluation against ground truth. To facilitate comparison with literature, synthetic SPR maps were also mapped to CT number using the established scanner-specific HLUTs and evaluated against real CTs. Quantitative evaluation was performed using Mean Absolute Error (MAE), Peak Signal-to-Noise Ratio (PSNR) and Structural Similarity Index Measure (SSIM) between real and synthetic images, defined as follows:

$$MAE = \frac{\sum_{i=1}^{N}|real(i) - syn(i)|}{N}$$

$$PSNR = 10\log_{10}\frac{L^2}{MSE}$$

$$SSIM = \frac{(2\mu_{real}\mu_{syn} + C_1)(2\delta_{real,syn} + C_2)}{(\mu_{real}^2 + \mu_{syn}^2 + C_1)(\delta_{real}^2 + \delta_{syn}^2 + C_2)}$$

With the following variables: $N$ = total number of voxels in the region of assessment, $L$ = dynamic range (2.6606 for SPR and 4095 for CT), MSE = mean squared error within the region of assessment, $\mu$ = the mean value of the image, $\delta^2$ = the variance of the image as indexed. $\delta_{real,syn}$ = the covariance of real and synthetic images. $C_1 = (k_1 L)^2$ and $C_2 = (k_2 L)^2$, and $k_1$ = 0.01 and $k_2$ = 0.03 (Wang *et al* 2004). SSIM was calculated for each pixel then averaged within the region of assessment (Wang *et al* 2004). Lower MAE suggests better pixel-wise accuracy, while higher PSNR and SSIM suggest better overall similarity of the synthetic images. Metrics were evaluated within the patient anatomy excluding the background by applying previously generated external masks. MAE was also evaluated for soft tissue, air, and bone separately, which were segmented on real CT by thresholding (air <-465 HU, bone >200 HU, and soft tissue ∈ [-465, 200] HU) (Nakano *et al* 2013).

**Dosimetric evaluation**

To evaluate the feasibility of using synthetic SPR maps for proton dose calculations, initial and adapted proton plans were retrospectively developed on an example patient with stage III (T4N2) HPV-associated squamous cell carcinoma of the tonsil by a proton dosimetrist. The tumor initially obstructed the airway but largely resolved mid-treatment which triggered clinical treatment adaptation based on mid-treatment CT simulation. Proton plans were developed in RayStation Treatment Planning System (v2024A SP3, Stockholm, Sweden) using pencil beam scanning techniques and a Hitachi ProBeat beam model. Each plan consisted of five beams including one anterior-posterior, two anterior obliques and two posterior obliques. Clinical contours were used to make the proton plans, where high-risk planning target volume (PTV), middle-risk clinical target volume (CTV) and low-risk CTV were prescribed to $D_{95}$=70, 60 and 54 Gy (relative biological effectiveness, RBE), respectively, in 33 fractions. Robust optimization was adopted including 3 mm setup uncertainty and 3.5% range uncertainty following standard clinical planning guidelines. Dose was calculated using Monte Carlo v5.6 dose engine with 0.5% statistical uncertainty. The plan was then recalculated on the CT images mapped back from CBCT-predicted SPR maps using the identical HLUT. To evaluate dosimetric differences, coverage (D95) of the PTV that received 70 Gy (RBE) and mean dose to OARs including parotids and submandibular glands were compared, and local 3D Gamma analysis was performed at 2 mm/2% and 3 mm/3% criteria with 10% dose threshold between the real and synthetic SPR dose using an in-house python code developed based on PyMedPhys.

To demonstrate the use of synthetic SPR model for APT, weekly CBCTs were collected to generate synthetic SPR maps for qualitative evaluation. Synthetic SPR maps from the same-day CBCT as the mid-treatment CT simulation were quantitatively evaluated against ground truth SPR maps derived from corresponding CT to facilitate interpretation of quantitative and dosimetric performance of the model.

**Additional comparative study**

Conventional deep learning methods were also implemented to provide a baseline for the performance of the proposed model. Unimodality residual network (ResNet) with and without perceptual loss and cGAN (pix2pix) (Isola *et al* 2017) with and without perceptual loss were tested. Five-fold cross validation was performed using the same data split, and data augmentation was adopted matching rBBrg. ResNet obtained the best quantitative and qualitative performance whereas cGAN-based results suffered from strong noise-like artifacts and ResNet with perceptual loss generated gridded artifacts in homogeneous tissue such as brain, thus ResNet was selected to report as the baseline. Specifically, a 12-block ResNet was trained to directly synthesize SPR from input CBCT/MRI, denoted as $G_R$ using the following training objective:

$$L = \lambda_R \| Y - G_R(X) \|$$

Where X is the input CBCT/MRI, Y is the paired ground truth SPR map, weight of the L1 norm $\lambda_R$ was set as 100 empirically. The model was optimized using Adam optimizer with learning rate of 1e-3, betas of 0.9 and 0.999, and eps of 1e-8. Model was quantitatively and dosimetrically evaluated following the same approach as rBBrg.

**Statistical analysis**

Non-parametric tests were performed given that no assumption was made on the distribution of the evaluated metrics. To compare the overall performance across multiple models, for each metric (MAE in different regions of assessment, PSNR and SSIM), an omnibus Kruskal-Wallis test was performed to detect statistically significant difference among the three models (modality-agnostic rBBrg, unimodality rBBrg and ResNet). P-values were adjusted applying Bonferroni correction for tests of multiple metrics, with P-values <0.05 considered statistically significant. When the Kruskal-Wallis test was significant, post-hoc pairwise comparisons were performed using Dunn's test to identify which pair(s) of models showed significant differences. Similarly, P-values were adjusted applying Bonferroni correction for tests of multiple pairs of models, with P-values <0.05 considered statistically significant.

## Results

***Table 1.*** *Summary of quantitative evaluation results on synthetic SPR from CBCT and MRI using different models with statistical test results. Metrics are reported in mean ± standard deviation [minimum, maximum]. P-values were adjusted using Bonferroni correction.*

| | Metric | Region | Modality-agnostic rBBrg | Uni-modality rBBrg | ResNet | Kruskal-Wallis P-value |
|---|---|---|---|---|---|---|
| **CBCT-based synthetic SPR results (n=37)** | MAE | External | 0.038 ± 0.004 [0.033, 0.047] | 0.038 ± 0.004 [0.032, 0.047] | 0.040 ± 0.004 [0.034, 0.053] | 0.05 |
| | | Soft tissue | 0.030 ± 0.003 [0.026, 0.039] | 0.030 ± 0.003 [0.026, 0.038] | 0.031 ± 0.003 [0.027, 0.042] | 1.00 |
| | | Bone | 0.059 ± 0.008 [0.046, 0.079] | 0.059 ± 0.008 [0.046, 0.080] | 0.067 ± 0.008 [0.051, 0.084] | <0.05 |
| | | Air | 0.127 ± 0.030 [0.086, 0.239] | 0.129 ± 0.030 [0.085, 0.244] | 0.140 ± 0.031 [0.091, 0.259] | 0.29 |
| | PSNR (d.B.) | External | 30.4 ± 0.8 [28.3, 31.7] | 30.4 ± 0.8 [28.3, 31.7] | 30.4 ± 0.8 [28.2, 31.8] | 1.00 |
| | SSIM (a.u.) | External | 0.94 ± 0.01 [0.91, 0.96] | 0.94 ± 0.01 [0.91, 0.96] | 0.93 ± 0.01 [0.90, 0.95] | 1.00 |
| **MRI-based synthetic SPR results (n=21)** | **Metric** | **Region** | **Modality-agnostic rBBrg** | **Uni-modality rBBrg** | **ResNet** | **Kruskal-Wallis P-value** |
| | MAE | External | 0.057 ± 0.005 [0.051, 0.068] | 0.058 ± 0.005 [0.051, 0.067] | 0.058 ± 0.006 [0.049, 0.072] | 1.00 |
| | | Soft tissue | 0.035 ± 0.004 [0.030, 0.045] | 0.036 ± 0.004 [0.030, 0.049] | 0.033 ± 0.003 [0.027, 0.042] | 0.13 |
| | | Bone | 0.134 ± 0.015 [0.113, 0.178] | 0.137 ± 0.014 [0.114, 0.169] | 0.140 ± 0.018 [0.106, 0.189] | 1.00 |
| | | Air | 0.184 ± 0.038 [0.133, 0.277] | 0.181 ± 0.040 [0.130, 0.266] | 0.209 ± 0.052 [0.132, 0.359] | 0.60 |
| | PSNR (d.B.) | External | 26.9 ± 0.8 [25.3, 28.1] | 26.8 ± 0.8 [25.5, 28.1] | 26.7 ± 1.0 [24.8, 28.7] | 1.00 |
| | SSIM (a.u.) | External | 0.88 ± 0.02 [0.84, 0.90] | 0.87 ± 0.02 [0.84, 0.90] | 0.87 ± 0.02 [0.83, 0.90] | 1.00 |

**CBCT-based SPR synthesis**

Table 1 summarizes key quantitative metrics evaluating the performance of the three models summarizing results for CBCT-based SPR synthesis. Modality-agnostic and unimodality rBBrg obtained comparable performance (P-value >0.05 for all tests). Krukal-wallis test found statistically significant difference in MAE within the bone, with pairwise post-hoc Dunn's test suggesting that modality-agnostic rBBrg outperformed ResNet (ΔMAE=11%, P-value <0.05) and unimodality rBBrg outperformed ResNet (ΔMAE=11%, P-value <0.05). PSNR and SSIM metrics were comparable across the models. To facilitate comparison with literature, predicted SPR values were mapped back to HU number using the corresponding HLUTs and evaluated against ground truth CT. The results are summarized in Supplement Table S1 with post-hoc Dunn test results summarized in Table S2. Similarly, modality-agnostic and unimodality rBBrg obtained comparable performance (MAE=52 HU within external, P-value >0.05 for all tests). Both rBBrg models outperformed ResNet results (Kruskal-Wallis and post-hoc Dunn test P-values <0.05 for MAE within the external (Δ~6-7%), MAE within the bone (Δ~12%) and SSIM within the external (Δ~3%)).

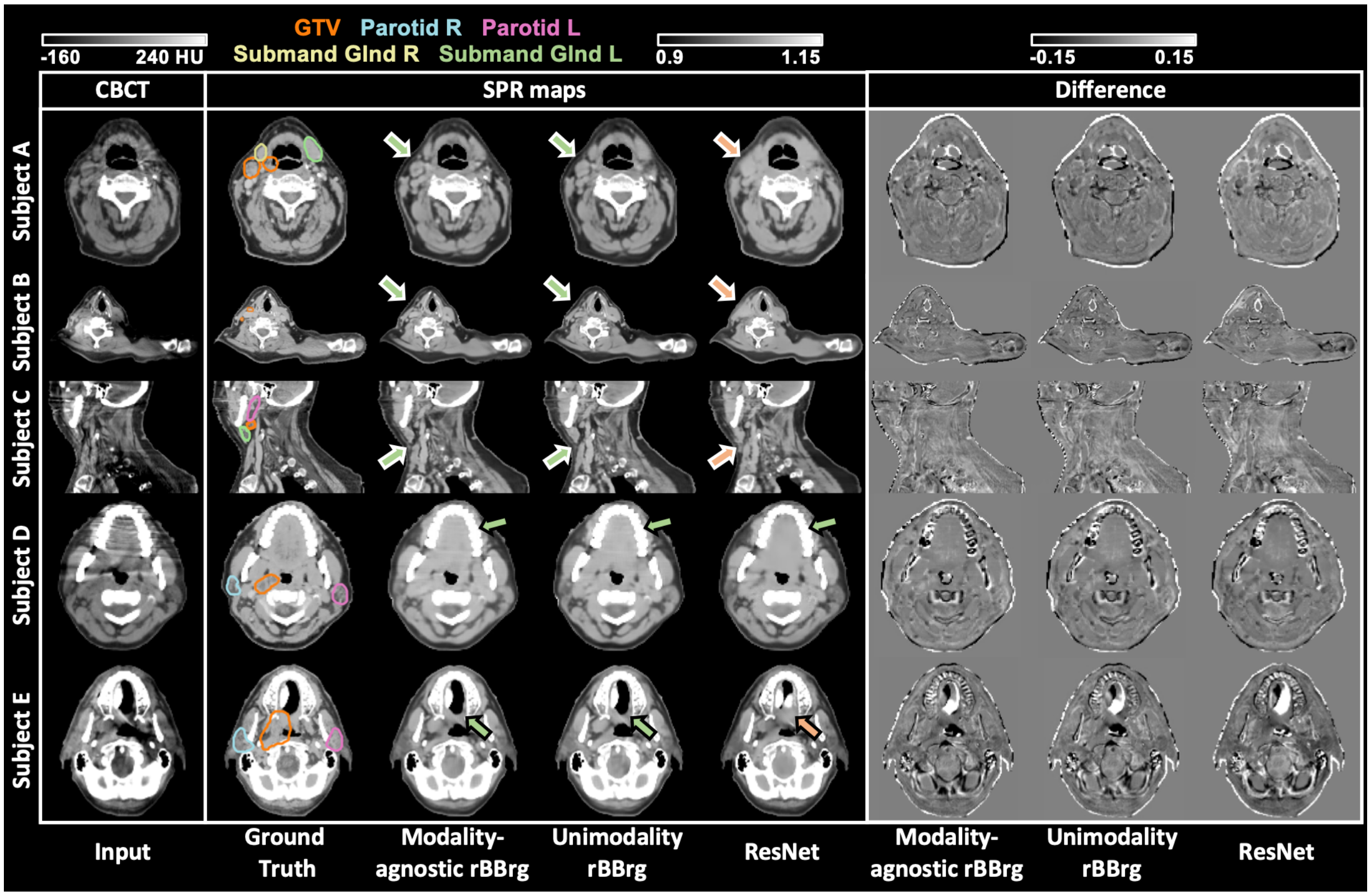

***Figure 3.*** *Key qualitative results of CBCT-based SPR synthesis for five representative subjects. Columns show CBCT, ground truth SPR maps, SPR map predictions of the three models including modality agnostic rBBrg, unimodality rBBrg, and ResNet, and corresponding difference maps relative to the ground truth. Green and orange arrows indicate regions of superior and inferior performance, respectively. Difference maps illustrated that modality-agnostic and unimodality rBBrg models demonstrated similar performance, both outperforming ResNet at preserving the anatomical fidelity.*

Figure 3 summarizes key qualitative results of CBCT-based SPR synthesis for five representative subjects. Modality-agnostic and unimodality rBBrg results showed comparable quality and excellent preservation of anatomical fidelity while ResNet generated more local errors. For Subject A, rBBrg models retained the tissue boundaries for the target and submandibular glands whereas ResNet generated blurred results. For Subject B, ResNet overestimated soft tissue abut to the target while rBBrg models faithfully predicted the anatomy. Despite being 2D models, rBBrg models obtained excellent tissue continuity along the slice direction as shown for Subject C. By contrast, ResNet results showed less smooth and less definitive soft tissue boundary with slight overestimation of soft tissue. For Subject D, the input CBCT demonstrated strong streaking artifacts. All models correctly predicted oral cavity anatomy with minimal artifacts. For Subject E, rBBrg models preserved the air-tissue interface in oral cavity adjacent to the target, while ResNet largely overestimated the soft tissue and generated hallucination of high-SPR structures in the oral cavity mimicking mouthpiece.

**MRI-based SPR synthesis**

Key quantitative metrics of MRI-based SPR synthesis are summarized in the lower half of Table 1. All three models showed comparable quantitative performance for all of the metrics except for marginal improvement in MAE within the air for modality-agnostic and unimodality rBBrg results compared to ResNet (12% and 13%, respectively). No statistically significant difference was found with Kruskal-Wallis test (P-value >0.05) for all metrics. Predicted SPR values were mapped to HU number and evaluated against ground truth CT with the results summarized in lower half of Supplement Table S1. Similarly, the three models showed comparable quantitative performance without statistically significant differences. MAE within external was around 84-85 HU for all three models.

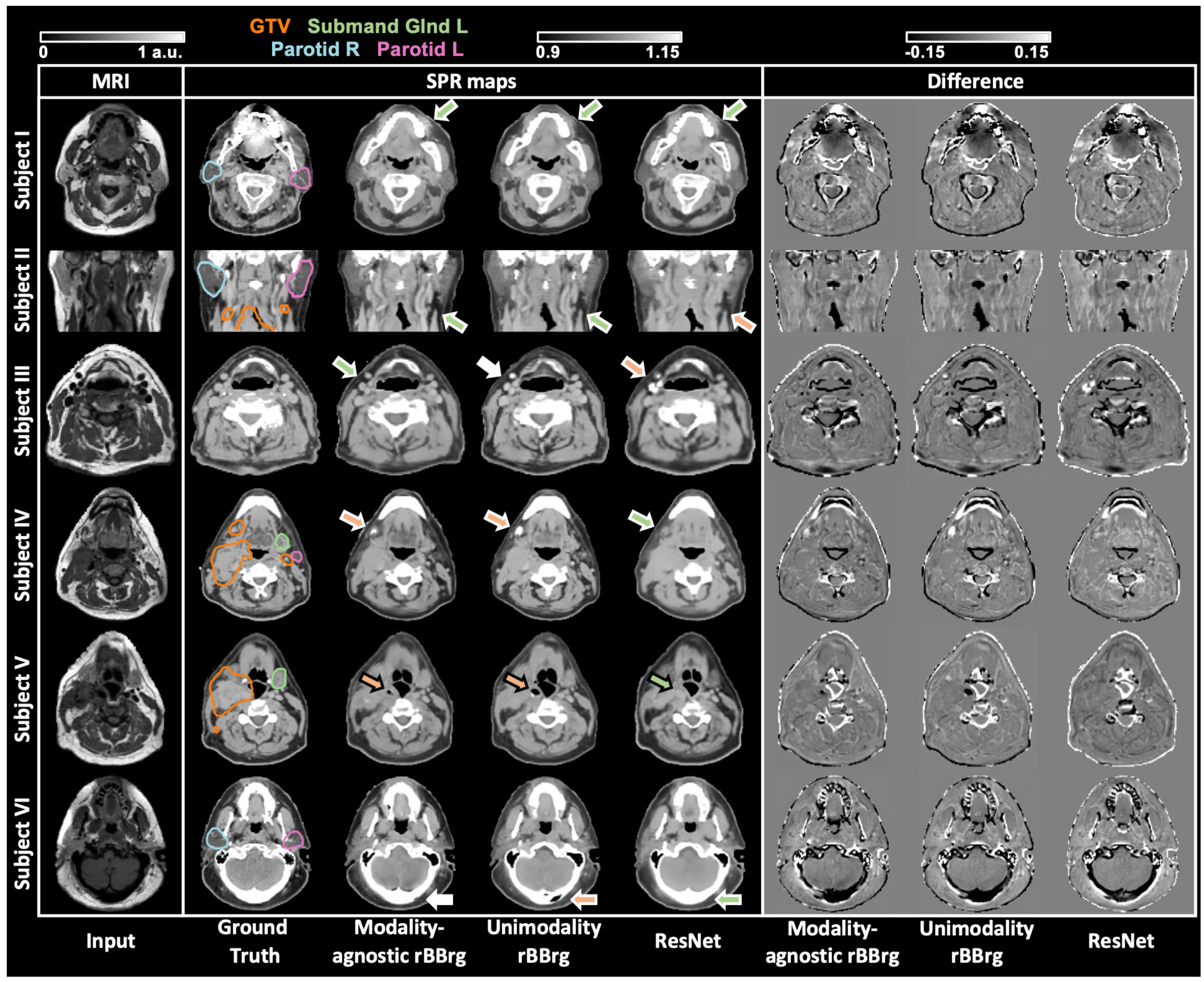

***Figure 4.*** *Key qualitative results of MRI-based SPR synthesis for six representative subjects. Columns show T1-weighted MRI, ground truth SPR maps, SPR map predictions of the three models including modality agnostic rBBrg, unimodality rBBrg, and ResNet, and corresponding difference maps relative to the ground truth. Green, white and orange arrows indicate regions of superior, intermediate and inferior performance, respectively. rBBrg models demonstrated locally better quality compared to ResNet.*

Figure 4 summarizes key qualitative results of MRI-based SPR synthesis for six representative subjects, demonstrating promising results with local challenges in MRI-based SPR synthesis. For Subject I, all three models predicted realistic dental structure and neighboring soft tissue without producing the streaking artifacts which appeared on the ground truth. Similar to CBCT-based results, all three models obtained excellent tissue continuity along the slice direction as shown for Subject II, while the two rBBrg models slightly better retained the target and soft tissue boundary. A key challenge presenting to all models was yield appropriate tissue classifications. For Subject III, several vessels showed low signal on the MRI, which were falsely predicted as high-SPR structures on ResNet results. Unimodality rBBrg showed slight misclassifications, whereas modality-agnostic model correctly predicted the vessels in this case. In addition, ResNet prediction appeared to be blur. For Subject IV, a small region of soft tissue adjacent to the mandible bone was incorrectly translated to high-SPR structure on both rBBrg model results while correctly predicted by ResNet. For Subject V, a vessel adjacent to the airway where anatomy was substantially deformed by the tumor was completely or partially predicted as air by rBBrg models. Note that ResNet showed similar issue for other subjects. Subject VI demonstrated rare misclassification of thick cortical bone in the skull as air on unimodality rBBrg model result.

## Case study: dosimetric evaluation

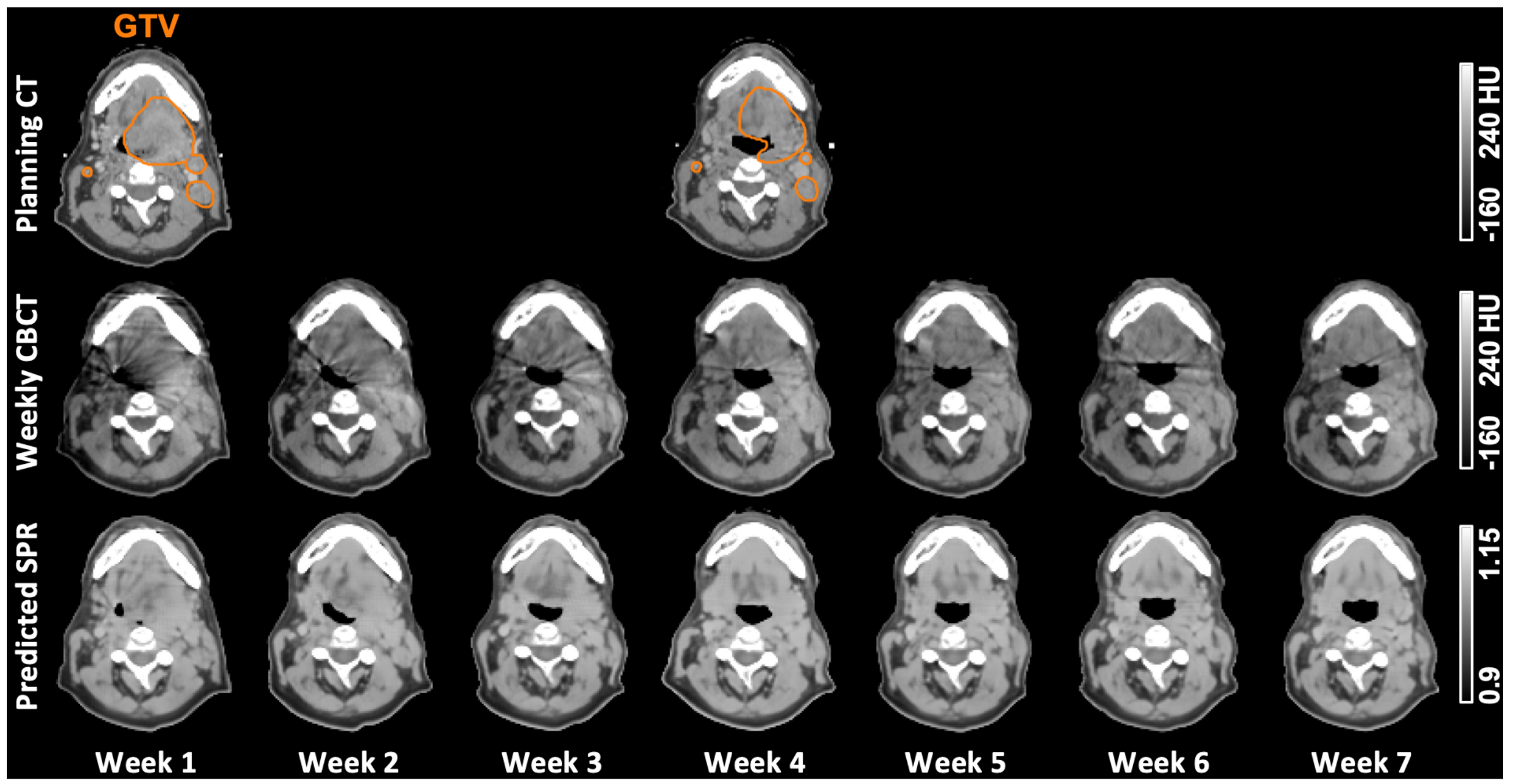

***Figure 5.*** *Planning CTs at pre-treatment and mid-treatment (week 4), weekly CBCT and corresponding sequential SPR maps predicted by the modality-agnostic rBBrg model for one example subject. Weekly CBCT suggested resolution of the tumor starting from week 2, prior to clinical mid-treatment adaptation. Sequential synthetic SPR maps demonstrated excellent quality and faithfully preserved the anatomy.*

Figure 5 demonstrated an example application of the synthetic SPR maps for APT. As shown in the initial planning CT and first-day CBCT, the tumor substantially obstructed the airway at treatment start. At week 4, clinical treatment adaptation was triggered due to tumor shrinkage, and the adapted plan was delivered 3 fractions after mid-treatment CT simulation. On the other hand, weekly CBCT suggested that the tumor shrinkage started since week 2, which could potentially necessitate adaptation if treated with proton. The sequential weekly SPR maps predicted by modality-agnostic rBBrg demonstrated excellent quality and faithfully preserved the anatomy including changing of the airway, suggesting the promise of synthetic SPR maps for rapid, accurate dosimetric evaluation to facilitate timely clinical decision-making in lieu of additional QACTs.

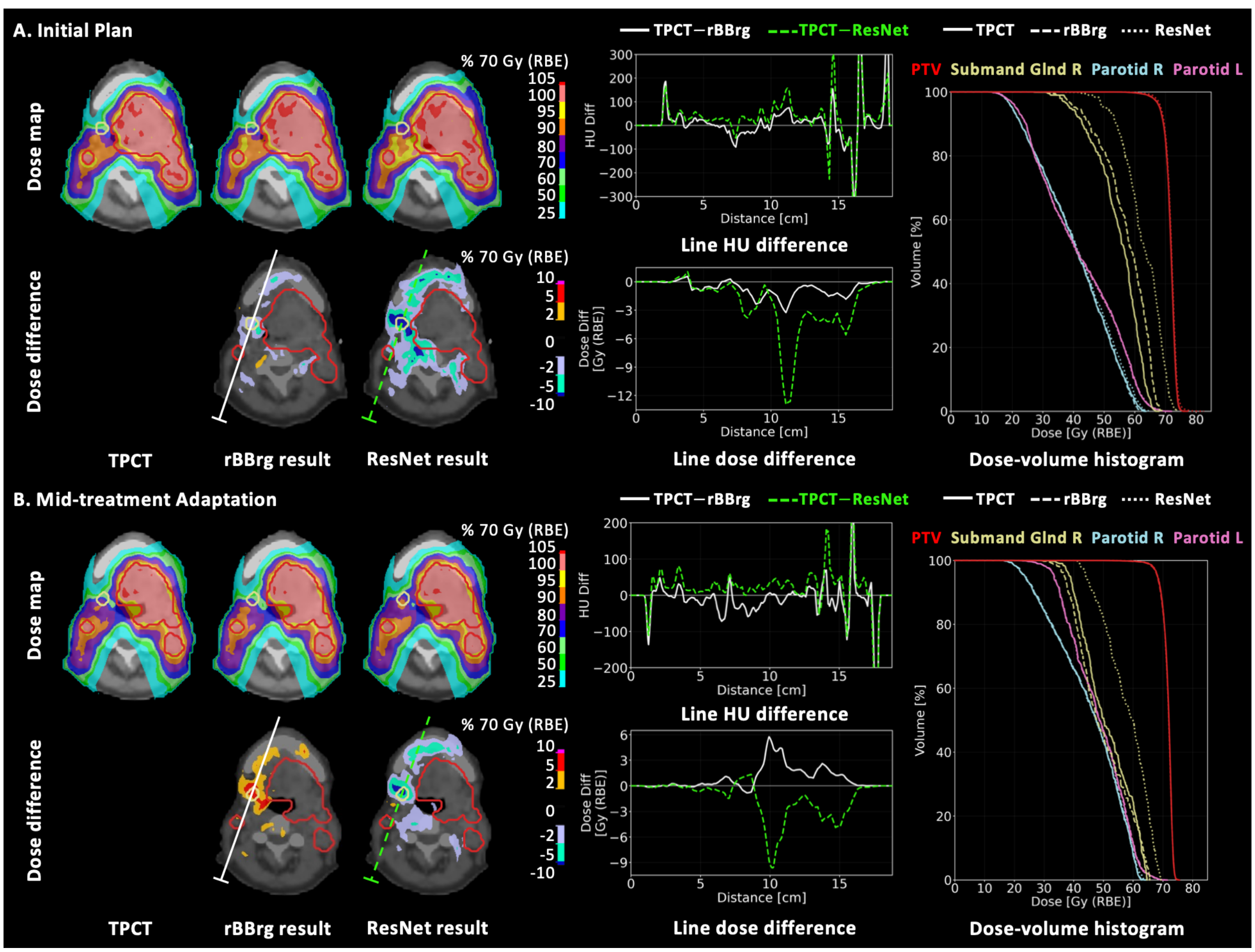

*Figure 6.* *Key dosimetric results of (A) initial plan and (B) mid-treatment adapted plan for the example patient. SPR maps were predicted using CBCT on (A) first-day and (B) same-day as mid-treatment CT, then mapped to HU number for dose calculation and evaluation. Modality-agnostic rBBrg prediction (note as "rBBrg") and ResNet prediction were evaluated, showing corresponding dose map, dose difference map to TPCT, and dose-volume histogram of key structures. A line going through the high dose difference region was drawn along one posterior oblique beam to extract line HU difference profile and line dose difference profile for assessment.*

Figure 6 demonstrated dosimetric evaluation of the example subject with key metrics summarized in Table 2. To evaluate dosimetric performance for initial and mid-treatment adapted plans, SPR maps were predicted using CBCT on first-day and same-day as mid-treatment CT, respectively, then mapped to HU number for dose calculation and dosimetric evaluation. Modality-agnostic rBBrg prediction and ResNet prediction were evaluated. Quantitatively rBBrg prediction obtained better performance with lower MAE within the external compared to ResNet prediction at both timepoints (Δ=-8/-12 HU for first-day/mid-treatment prediction, respectively). As a result, for both initial and mid-treatment adapted plans, rBBrg achieved better dosimetric performance indicated by higher gamma pass rates when evaluating against ground truth dose (4.3%/2.4% higher compared to ResNet results at 2 mm/2 % for initial/mid-treatment plans, respectively). Nevertheless, both model prediction achieved gamma pass rate >90% at both timepoints. All predictions demonstrated excellent agreement in DVH metrics including PTV $D_{95}$ and $D_{mean}$ for both parotids (≤1% for both models at both timepoints), except for $D_{mean}$ for right submandibular gland which fell in the high dose gradient region at the distal end of two beams, resulting in large discrepancies (maximum of 14.8% in $D_{mean}$ comparing initial plan ground truth and ResNet prediction). DVH curves led to the same observations as the metrics. For both plans, curves for PTV and both parotids showed excellent agreement, whereas the curves for right submandibular gland showed substantial discrepancies, which were more substantial in ResNet results. A line was drawn along the direction of the posterior oblique beam. For both timepoints, line HU difference profiles suggested a sharp spike (~100-200 HU) at the beam entrance potentially due to external mismatch. Line HU profiles also demonstrated systematic underestimation of HU in soft tissue for ResNet results leading to constant build-up of range errors, whereas HU differences in rBBrg results oscillated more around 0, suggesting potential cancellation in range error accumulation. Line dose difference profile demonstrated substantially higher dose discrepancies in ResNet prediction at both timepoints as demonstrated in Figure 6.

***Table 2.*** *Key metrics evaluating quantitative and dosimetric performance of CBCT-based modality-agnostic rBBrg prediction (note as "rBBrg") and ResNet prediction for the example subject at the two timepoints, including mean absolute error (MAE) within the external, gamma pass rates, and dose-volume histogram (DVH) metrics (reported in absolute and percentage differences).*

| | **Metrics** | | **rBBrg** versus **TPCT** | **ResNet** versus **TPCT** |
|---|---|---|---|---|
| **Initial plan** | MAE within external | SPR | 0.037 | 0.042 |
| | | HU | 48 | 56 |
| | Gamma pass rate | 2 mm/2 % | 98.8% | 94.5% |
| | | 3 mm/3 % | 99.7% | 97.4% |
| | DVH metric difference (absolute Gy [RBE]/%) | PTV $D_{95}$ | 0.0 (0.0%) | 0.4 (0.6%) |
| | | Submandibular gland R $D_{mean}$ | 2.5 (4.6%) | 8.1 (14.8%) |
| | | Parotid R $D_{mean}$ | 0.0 (0.1%) | 0.4 (1.0%) |
| | | Parotid L $D_{mean}$ | 0.0 (0.0%) | -0.1 (-0.2%) |
| **Mid-treatment adaptation** | MAE within external | in SPR | 0.036 | 0.047 |
| | | In HU | 46 | 58 |
| | Gamma pass rate | 2 mm/2 % | 99.3% | 96.9% |
| | | 3 mm/3 % | 99.8% | 98.8% |
| | DVH metric difference (absolute Gy [RBE]/%) | PTV $D_{95}$ | -0.1 (-0.2%) | 0.0 (0.1%) |
| | | Submandibular gland R $D_{mean}$ | -1.4 (-2.7%) | 6.9 (13.5%) |
| | | Parotid R $D_{mean}$ | 0.1 (0.3%) | 0.1 (0.2%) |
| | | Parotid L $D_{mean}$ | 0.1 (0.2%) | -0.1 (-0.2%) |

## Discussion

A novel modality-agnostic regularized Brownian bridge diffusion model (rBBrg) was developed and validated for rapid, deterministic prediction of SPR maps from auxiliary RT datasets, including CBCT or T1-weighted MRI in our experiments, to support APT. For both CBCT and MRI-based SPR map prediction, modality-agnostic and unimodality rBBrg demonstrated comparable performance. Both rBBrg models outperformed ResNet at CBCT-based prediction, quantitatively in several metrics and qualitatively in preserving the anatomical fidelity, whereas for MRI-based prediction, the three models obtained comparable global performance.

For CBCT-based SPR prediction, modality-agnostic rBBrg faithfully preserved patient anatomy with excellent quantitative performance (MAE=0.038 in SPR/52 HU within external). While studies on SPR prediction from CBCT are currently limited, our results are comparable to literature which reported MAE=0.06 (Harms *et al* 2020). Evaluating in HU, our results are also comparable to literature reporting MAE of CBCT-based HN CT synthesis within the range of ~20-80 HU (Altalib *et al* 2025). Comparing across models, modality-agnostic and unimodality rBBrg demonstrated similar qualitative and quantitative performance (P-values >0.05 for all statistical tests) while both outperformed baseline ResNet (ΔMAE~6-7% evaluating in HU, P-values <0.05 for Kruskal-Wallis test and post-hoc pairwise Dunn's test), which demonstrated inferior anatomical fidelity in qualitative results such as loss of soft tissue boundary and overestimation of soft tissue. While dosimetric evaluation of two proton plans on a representative case suggested local discrepancies (>5 Gy [RBE]) in high gradient region at the distal end of beam, excellent overall agreement was found between dose calculation on modality-agnostic rBBrg prediction and ground truth CT (PTV70 $|\Delta D_{95}|<0.1\%$, gamma pass rate≥98.8% at 2 mm/2 %), outperforming ResNet results (gamma pass rate≤96.9% at 2 mm/2 %), suggesting promise of applying rBBrg for accurate proton dose calculation in support of APT.

For MRI-based SPR prediction, the three models obtained comparable quantitative performance with MAE~0.058 in SPR/85 HU within the external contour (Kruskal-Wallis P-value >0.05 for all metrics). At present, similar studies on SPR prediction from MRI are limited (Wang *et al* 2022). Modality-agnostic rBBrg obtained MAE of a similar magnitude to literature (Wang *et al* 2022) which evaluated predicted SPR after conversion to HU number and reported MAE=42 HU in pediatric brain. Evaluated in HU, our results are also comparable to the literature reporting MRI-based synthetic CT with MAE of 60-100 HU (Boulanger *et al* 2021). Qualitatively, prediction was primarily challenged in classification of tissue with low MRI signal, such as bone, air and vessels in our dataset, which is an inherent challenge with MRI datasets and commonly reported in MRI-based synthetic CT literature (Singhrao *et al* 2023). The choice of the MRI sequence in our experiment was based on the image quality and available clinical MR protocol. Literature have reported to incorporate multiple clinical MRI sequences as model input to potentially improve tissue classification (Tie *et al* 2020), or incorporate specific sequences such as zero TE (ZTE) MRI which provides bone signal to improve the bone prediction performance (Lauwers *et al* 2025) but remains less standard in radiotherapy setting (Wiesinger and Ho 2022) and have yet to be explored for proton therapy.

In our experiments, modality-agnostic rBBrg achieved comparable performance to unimodality rBBrg, despite the added complexity in the modality-agnostic approach of training a single model to learn SPR synthesis from two distinct input modalities. Through the use of class embedding, modality-agnostic rBBrg only requires a single U-net for inference, whereas unimodality approach requires two separate U-nets for CBCT-SPR and T1-SPR synthesis, respectively. Moreover, modality-agnostic approaches can readily scale to include additional imaging modalities such as other MRI contrasts, DECT, or vertical CT, offering greater advantage in efficiency as the number of modalities increases. Enabling modality-agnostic SPR prediction offers more flexibility for clinical implementation, while studies on such approach have been limited especially in the context of supporting adaptive radiation therapy (Wongtrakool *et al* 2025).

During collection of paired MRI-CT data, it was found that many patients were applied with oral immobilization devices (e.g., mouthpieces in the form of bite blocks (Carbon® Cyanate Ester-based resin with physical density ~1.20 g/cm$^3$) or dental impression putty blocks (Correct Plus™ vinyl polysiloxane with physical density ~1.46 g/cm$^3$)) during CT simulation to displace normal tissue from high dose regions such as tongue and soft palate (Singh *et al* 2021). However, these mouthpieces were not applied during MR simulation which may introduce image quality degradations such as introducing signal voids (Schmidt and Payne 2015) as well as patient discomfort. This mismatch and resultant differences in jaw position and oral cavity confounded creation of paired dataset. MRI-CT mismatch due to inconsistent application of oral immobilization devices has been reported elsewhere and was noted to impact the evaluation of their results (Singhrao *et al* 2023), thus such data including oral immobilization devices were not included in our MRI cohort. This remains a practical challenge as synthetic SPR maps predicted from MRI acquired without mouthpiece do not necessarily reflect patient anatomy in treatment position, which is beyond the scope of our study.

A few other limitations of our study lie in the process of creating the paired dataset by deformably registering CBCT/MRI to SECT, followed by calibrating SECT to SPR. Potential registration uncertainties exist which might confound the model training and evaluation. In addition, although iMAR algorithm was adopted to reduce dental artifacts, residual artifacts remained on the CT for some patients. Furthermore, the mismatch of air in oral cavity sometimes could not be addressed by DIR. These factors combined result in less accurate ground truth for model training and evaluation. Moreover, ground truth SPR was derived from SECT. Many studies have showed that dual energy CT (DECT) provides better SPR estimations, which reduced range uncertainties from 3.5% to 2.2% compared to SECT-based stoichiometric methods (Li *et al* 2017, Bär *et al* 2018). However, clinical implementation of DECT for SPR estimations is currently very limited, challenged by motion in the disease sites, lack of experience, etc. (Peters *et al* 2023, Taasti *et al* 2018). While an example adaptive HN patient was presented, dosimetric evaluation in larger cohort is warranted for further evaluation of SPR prediction performance.

## Conclusions

A novel modality-agnostic rBBrg model was trained to synthesize SPR maps from CBCT or T1-weighted MRI datasets. The framework allowed rapid and deterministic inference while leveraging the diffusion formulation. Quantitative evaluation suggested that modality-agnostic rBBrg was comparable to unimodality rBBrg, superior to the conventional ResNet method for CBCT-based prediction, and comparable to ResNet for MRI-based prediction. Preliminary dosimetric evaluation suggested the promise of rBBrg for precise proton dose calculation. With further dosimetric validation in larger cohorts, rBBrg is promising for providing accurate SPR maps to facilitate rapid APT.

## Supplementary Materials

Supplementary materials for Brownian bridge methodology and additional results for model evaluation in HU

### S1. Detailed expression of discretized reverse process of Brownian bridge

For discrete time step with a total number of time steps T, note $\mathbf{X}(t=0)=\mathbf{X}_0, \mathbf{X}(t=\mathrm{T})=\mathbf{X}_\mathrm{T}, \boldsymbol{\epsilon} \sim \mathcal{N}(0, \mathbf{I})$, $\boldsymbol{\epsilon}_{\boldsymbol{\theta}}$ representing the Brownian bridge model that learns the difference between $\mathbf{X}_t$ and $\mathbf{X}_0$, then the reverse process can be discretized as:

$$\mathbf{X}_{t-1} = C_{xt}\mathbf{X}_\mathrm{t} - C_{yt}\mathbf{X}_\mathrm{T} - C_{\epsilon t}\boldsymbol{\epsilon}_{\boldsymbol{\theta}}(\mathbf{X}_t, t) - C_{zt}\boldsymbol{z}, \quad t = T, \dots, 2$$

where $\boldsymbol{z} \sim \mathcal{N}(0, \mathbf{I})$, and

$$C_{xt} = \frac{1}{C}\left(1 + \frac{1}{\mathrm{T}\ln\frac{\mathrm{T}}{\mathrm{T}-t+1}}\right)$$

$$C_{yt} = \frac{1}{C}\left(\frac{1}{\mathrm{T}-t+1} - \frac{t-1}{\mathrm{T}\,(\mathrm{T}-t+1)\ln\frac{\mathrm{T}}{\mathrm{T}-t+1}}\right)$$

$$C_{\epsilon t} = \frac{1}{C}\frac{1}{\mathrm{T}\ln\frac{\mathrm{T}}{\mathrm{T}-t+1}}$$

$$C_{zt} = \frac{\lambda}{C}\sqrt{\frac{\mathrm{T}-t+1}{\mathrm{T}}}\sqrt{\frac{1}{\mathrm{T}}}$$

$$C = 1 - \frac{1}{\mathrm{T}-t+1} + \frac{1}{(\mathrm{T}-t+1)\ln\frac{\mathrm{T}}{\mathrm{T}-t+1}})$$

For t=1, $\mathbf{X}_0 = \mathbf{X}_1 - \boldsymbol{\epsilon}_{\boldsymbol{\theta}}(\mathbf{X}_1, 1)$.

**Table S1.** Summary of CT number-based quantitative metrics on synthetic SPR from CBCT and MRI using different models with statistical test results. Synthetic SPR maps were mapped back to HU numbers using the established Hounsfield look-up tables, then evaluated against ground truth CT in HU number. Metrics are reported in mean ± standard deviation [minimum, maximum]. P-values were adjusted using Bonferroni correction.

| | Metric | Region | Modality-agnostic rBBrg | Uni-modality rBBrg | ResNet | Kruskal-Wallis P-value |
|---|---|---|---|---|---|---|
| **CBCT-based synthetic SPR results in HU (n=37)** | MAE (HU) | External | 52 ± 6 [44, 65] | 52 ± 6 [43, 64] | 55 ± 6 [45, 75] | <0.05 |
| | | Soft tissue | 38 ± 4 [33, 49] | 38 ± 4 [33, 48] | 40 ± 4 [34, 54] | 1.00 |
| | | Bone | 118 ± 17 [93, 161] | 118 ± 17 [93, 161] | 133 ± 17 [103, 172] | <0.05 |
| | | Air | 126 ± 30 [85, 234] | 128 ± 30 [84, 239] | 137 ± 31 [88, 251] | 0.35 |
| | PSNR (d.B.) | External | 32.2 ± 1.1 [29.6, 34.1] | 32.2 ± 1.1 [29.5, 34.0] | 31.9 ± 1.1 [29.5, 34.0] | 1.00 |
| | SSIM (a.u.) | External | 0.89 ± 0.02 [0.83, 0.93] | 0.89 ± 0.03 [0.82, 0.93] | 0.86 ± 0.03 [0.80, 0.92] | <0.05 |
| | **Metric** | **Region** | **Modality-agnostic rBBrg** | **Uni-modality rBBrg** | **ResNet** | **Kruskal-Wallis P-value** |
| **MRI-based synthetic SPR results in HU (n=21)** | MAE (HU) | External | 84 ± 8 [73, 102] | 85 ± 8 [74, 105] | 85 ± 10 [72, 110] | 1.00 |
| | | Soft tissue | 48 ± 5 [39, 58] | 49 ± 6 [39, 65] | 45 ± 5 [36, 59] | 0.43 |
| | | Bone | 257 ± 27 [218, 327] | 262 ± 27 [216, 316] | 269 ± 34 [215, 347] | 1.00 |
| | | Air | 186 ± 39 [134, 277] | 182 ± 40 [131, 272] | 209 ± 52 [132, 357] | 0.81 |
| | PSNR (d.B.) | External | 27.3 ± 1.0 [25.4, 28.8] | 27.2 ± 0.9 [25.6, 28.7] | 27.2 ± 1.1 [24.8, 28.7] | 1.00 |
| | SSIM (a.u.) | External | 0.82 ± 0.02 [0.77, 0.85] | 0.82 ± 0.02 [0.78, 0.85] | 0.81 ± 0.03 [0.74, 0.86] | 1.00 |

**Table S2.** Summary of post-hoc Dunn test results for metrics that showed significant Kruskal-Wallis test results comparing synthetic SPR (evaluated in HU) from CBCT and MRI across the three models.

<table>
<tr><td rowspan="4">CBCT-based synthetic SPR results in HU (n=37)</td><td>Metric</td><td>Region</td><td>Modality-agnostic rBBrg<br>versus<br>Unimodality rBBrg</td><td>Modality-agnostic rBBrg<br>versus<br>ResNet</td><td>Uni-modality rBBrg<br>versus<br>ResNet</td></tr>
<tr><td>MAE (HU)</td><td>External</td><td>1.00</td><td><0.05</td><td><0.05</td></tr>
<tr><td>MAE (HU)</td><td>Bone</td><td>1.00</td><td><0.05</td><td><0.05</td></tr>
<tr><td>SSIM (a.u.)</td><td>External</td><td>1.00</td><td><0.05</td><td><0.05</td></tr>
</table>